\documentclass[aps,prl,twocolumn,superscriptaddress,floatfix,letter,10pt]{revtex4-1}
\usepackage[pdfpagemode=None,pdfstartview=FitH]{hyperref}
\usepackage{hyperref}
\usepackage{graphicx}
\usepackage{epstopdf}
\usepackage{mathptmx}
\usepackage[scaled=0.90]{helvet} % ss
\usepackage{courier} % tt
\normalfont
\usepackage[T1]{fontenc}
\usepackage{amsmath,amssymb,amsthm}

\DeclareMathOperator{\Tr}{Tr}

\begin{document}
\title{The microscopic origin of thermodynamic entropy in isolated systems}

\author{J. M. Deutsch}
\affiliation{Department of Physics, University of California, Santa Cruz CA 95064}
\author{Haibin Li}
\affiliation{Department of Physics, University of California, Santa Cruz CA 95064 and 
Department of Applied Physics, Zhejiang University of Technology, Hangzhou
310023, P. R. China}
\author{Auditya Sharma}
\affiliation{International Institute of Physics, Federal University of Rio Grande do Norte, Brazil}

\begin{abstract}
A microscopic understanding of the thermodynamic entropy in quantum
systems has been a mystery ever since the invention of quantum
mechanics. In classical physics, this entropy is believed to be the
logarithm of the volume of phase space accessible to an isolated
system~\cite{MaStatMech}.  There is no quantum mechanical analog
to this. Instead, Von Neumann's hypothesis for the entropy~\cite{VonNeumann}
is most widely used. However this gives zero for systems with a known
wave function, that is a pure state.  This is because it measures
the lack of information about the system rather than the flow of heat
as obtained from thermodynamic experiments. Many arguments attempt to
sidestep these issues by considering the system of interest coupled
to a large external one, unlike the classical case where Boltzmann's
approach for isolated systems is far more satisfactory. With
new experimental techniques, probing the quantum nature of
thermalization is now possible~\cite{Kinoshita,Hofferberth}. Here,
using recent advances in our understanding of quantum
thermalization~\cite{DeutschQuantStatMech,SrednickiQuantStatMech,RigolDunjkoOlshanii,SantosRigol,RigolBreakdown,RigolQuantumQuench}
we show how to obtain the entropy as is measured from thermodynamic
experiments, solely from the self-entanglement of the wavefunction, and find strong numerical evidence
that the two are in
agreement for non-integrable systems. It is striking that this entropy,
which is closely related to the concept of heat, and generally thought
of as microscopic chaotic motion, can be determined for systems in energy
eigenstates which are stationary in time and therefore not chaotic, but 
instead have a very complex spatial dependence.
\end{abstract}

\maketitle

The emergence of thermodynamics from underlying microscopic motion is
still unclear.  One of the hardest concepts to understand is that of
entropy. 
Classically, the microscopic
state of a system can be thought of as a single point in phase space and
as it evolves, this point wanders in this space, filling up more of it
as time progresses.  Boltzmann's hypothesis relates the entropy to the
volume that can be filled, and shows how microscopic motion determines
large scale thermodynamic properties~\cite{MaStatMech}. For a generic,
that is a non-integrable, system, ergodicity implies that this volume
will be the hypersurface of constant energy, but for integrable systems,
the volume accessible depends on initial conditions so that such systems
will not thermalize.

There are many different approaches and definitions for understanding
entropy~\cite{Balian}, but here we will be concerned with the experimental
thermodynamic definition that is obtained, for example, by doing
calorimetry measurements as related to the flow of heat. Classically,
these thermodynamic measurements are not influenced by knowledge of the
microscopic state of a system. For example, computer simulations are
often used to predict thermodynamic quantities, such as entropy. However
microscopic knowledge of the system allows one to extract work out of
it by the well known example of Maxwell's demon. This would violate the
second law of thermodynamics unless we correspondingly reduce the entropy
to take into account this microscopic information~\cite{Balian}. Therefore
classically there are two distinct uses of the word entropy: ({\em i})
obtained by performing macroscopic thermodynamic measurements, which we will call the
thermodynamic entropy, and ({\em ii}) a measure of the lack of available
information about a system's state.

The quantum mechanical situation is far less clear. Knowledge of the
system through observation does alter the state of the system, so that it
is not apparent if the thermodynamic entropy is altered by this process.
For a system with density matrix $\rho$, the Von Neumann entropy  $S_{VN}
= - \Tr (\rho \ln \rho)$ which is identically {\em zero} for any system
in a pure state, including an energy eigenstate. Because the state of
the system is known completely, this reduces its entropy  to zero as
in case ({\em ii}). But what does this do to entropy ({\em i})? If we
do thermodynamic measurements on systems in pure states, that is, where
the state is known completely to the experimenter, does this alter what is measured?

Furthermore, the quantum ensemble definition of entropy, case ({\em i}), is
simply related to the density of states of the energy. But if the system
is in a single energy eigenstate, it might appear to be impossible to
get the density of states. This would argue that thermodynamic
entropy measured for systems in pure states, would be different
than for a system coupled to an external environment.  Such questions
until recently were purely theoretical but now with the emergence of
experiments that study thermalization of isolated quantum systems or
lack thereof~\cite{Kinoshita,Hofferberth}, a proper understanding of
the microscopic origins of entropy has become increasingly important.

Here we find convincing evidence that for pure states, there is a way of defining an
entropy as in case ({\em i}) for a non-integrable quantum mechanical
system with a large number of degrees of freedoms which is identical
to the thermodynamic definition and give an explicit prescription
for its construction from a knowledge of the eigenstate alone, thus
successfully linking the thermodynamic entropy with its underlying
microscopic origins. Throughout this work we will be concerned with
the thermodynamic limit of a large number of degrees of freedom, $N$, and
will not consider non-extensive corrections to the entropy for finite
systems. Furthermore in this limit, the microcanonical and canonical
ensembles give identical results~\cite{MaStatMech}.  Therefore we can
freely choose which ensemble to consider.

Statistical mechanics is concerned mainly with computing the time average
of observables that depend on only a few variables, and relates this time
average to an average over an ensemble. 
Thus for a system with average energy $\langle E\rangle$, statistical mechanics posits
the relation between the time average of some observable $\hat O$ and its average
over the microcanonical ensemble of states, the latter
being far easier to compute:
\begin{equation}
\label{eq:StatMech}
\overline {\langle {\hat O}\rangle} = \Tr (\rho_{micro,\langle E\rangle} {\hat O}) 
\end{equation}
where the bar on the left hand side denotes a time average and the microcanonical density matrix at energy $E_0$ is
\begin{equation}
\label{eq:rhomicro}
\rho_{micro,E_0} \equiv \frac{1}{{\cal N}_{states}} \sum_{E_0 < E < E_0+\Delta E} |E\rangle \langle E|
\end{equation}
and $\Delta E$ is much greater than the average distance between
neighboring energy levels but much less than the macroscopic energy scale
$E_0$, and ${\cal N}_{states}$ is the number of terms in the sum. In quantum
mechanics, the fluctuation in the energy can be
large~\cite{DeutschQuantStatMech} but in practice is
taken to be small so that $\langle E\rangle$ is a good measure of the energy. The way
to see the connection between the underlying quantum evolution needed on
the left hand side of Eq. \ref{eq:StatMech} of a generic system  and the
ensemble methods of statistical mechanics has recently become much better
understood.  The idea is the ``Eigenstate Thermalization Hypothesis"
(ETH)~\cite{DeutschQuantStatMech,SrednickiQuantStatMech,RigolDunjkoOlshanii}:
for large $N$, the expectation value of an observable in an energy
eigenstate becomes equal to the microcanonical average at the same energy,
that is
\begin{equation}
\label{eq:ETH}
\Tr (\rho_E {\hat O}) = \Tr (\rho_{micro,E} {\hat O})
\end{equation}
where $\rho$ is the density matrix for the wavefunction
$|E\rangle$ at energy $E$, $\rho_E = |E\rangle\langle E|$. Put
more simply, the expectation value of $\hat O$ will vary very
little between neighboring energy levels for large $N$, implying
that the expectation value of $\hat O$ in any energy eigenstate
is the microcanonical average (at that energy).  There is strong
numerical evidence and analytic arguments to support this for a variety of
systems~\cite{DeutschQuantStatMech,SrednickiQuantStatMech,RigolDunjkoOlshanii,SantosRigol,RigolBreakdown,RigolQuantumQuench}.
Not all systems obey this hypothesis~\cite{Basko,OganesyanHuse}. In particular,
integrable systems, do not do so. Note that $\rho_E \ne
\rho_{micro,E}$, as there a large number of constants of motion, even for
generic Hamiltonians, such as projection operators\cite{Sutherland} that
violate Eq. \ref{eq:ETH}. This is why the choice of observables
satisfying this equation is restricted.  

Below we shall extend these
ideas to understanding entropy.  In the thermodynamic limit statistical
mechanics posits the formula for the thermodynamic entropy $S_{thermo}
= -\Tr (\rho_{micro} \ln \rho_{micro})$ where $\rho_{micro}$  could be
equivalently replaced with the thermal density matrix in the canonical
ensemble in the limit that we are considering. We want to see if this
can be calculated from the properties of the wavefunction itself without
any recourse to ensembles.

We consider homogeneous systems for simplicity with short range
interactions and start by following the standard textbook scenario: the
system is divided into two parts, the larger part, $B$, can act as a bath
in contact with the smaller one, $A$, that is our system of interest. If
the complete system starts out in a pure state, then by doing a partial
trace over $B$, the reduced density matrix for $A$,  $\rho_A = \Tr_B
\rho$ becomes mixed because it is entangled with $B$. This is part of
a common but non-rigorous argument for how a canonical distribution of
energies emerge for the smaller system $A$, where the temperature $T$
is given by the statistical mechanical relation between $T$ and the
average energy. If the complete system was actually integrable this
argument fails and  thermalization does not occur.

The {\em entanglement entropy} is defined as
\begin{equation}
\label{eq:S_Ent}
S_{Ent}(A,B) = -\Tr (\rho_A \ln \rho_A). 
\end{equation}
which has been studied in detail for a wide variety of systems
and is very important in the study of quantum information
theory\cite{BennetQuantInfo}, black holes~\cite{Bekenstein} and quantum
phase transitions~\cite{Vidal}. This will also be important in what
follows.

If  instead we considered the complete system to already be described by the
microcanonical density matrix, then $\rho_{A,micro} \equiv \Tr_B
\rho_{micro}$ must behave with canonical statistics even for integrable
systems, and the entanglement entropy in that case is $S_{Ent,micro}(A,B) \equiv
-\Tr (\rho_{A,micro} \ln \rho_{A,micro})$ which is the statistical
mechanical expression for the entropy. This entropy is what one measures
experimentally in thermodynamic experiments. Thus
\begin{equation}
\label{eq:Smicro=thermo}
S_{thermo}(A) = S_{Ent,micro}(A,B)  = -\Tr (\rho_{A,micro} \ln\rho_{A,micro})
\end{equation}
This argument is very similar to most treatments of entropy in
textbooks\cite{LandauStatMech} however there is more evidence than
that this to support it. Rigorous analysis~\cite{CalabreseCardyReview}
gives an canonical entanglement entropy for one dimensional models in
agreement with the thermodynamic results~\cite{Korepin}.

This {\em does not} connect thermodynamic entropy with the wavefunction
of a system because the right hand side of Eq. \ref{eq:Smicro=thermo}
involves the microcanonical average  and is similar to the right hand side
of the ETH as written in Eq. \ref{eq:ETH}, but differs because of the
nonlinear logarithmic factor which is not an observable operator. Thus
we ask whether we can extend the ETH to quantities of this form; that
is, for a generic quantum system in an energy eigenstate can we replace
$\rho_{A,micro}$ with $\rho_{A}$? In other words is
\begin{equation}
\label{eq:Sthermo=rhologrho}
S_{thermo}(A) = - \Tr (\rho_A \ln \rho_A) ?
\end{equation}
This form was recently hypothesized using other less direct analytical
arguments~\cite{DeutschEntropy}. Although this is the thermodynamic entropy
for a subsystem of $A+B$, the entropy is extensive so we can obtain the entropy
for the full system by adding the entropies for individual subsystems together.
In the particular case of a homogeneous system, the above equation can easily give us the
thermodynamic entropy per unit volume. Thus entropy defined this way is a
measure of {\em self-entanglement}.

Now we employ exact diagonalization on a number of systems to determine if Eq.
\ref{eq:Sthermo=rhologrho} is satisfied, and how it scales with system size.
Because system sizes are still far from the thermodynamic limit we will minimize
finite size effects by asking equivalently using Eq. \ref{eq:Smicro=thermo}, does
\begin{equation}
\label{eq:rho_micro=rho}
 -\Tr (\rho_A \ln\rho_A) = -\Tr (\rho_{A,micro} \ln\rho_{A,micro})?
\end{equation}

\begin{figure}[htp]
\begin{center}
(a)\includegraphics[width=\hsize]{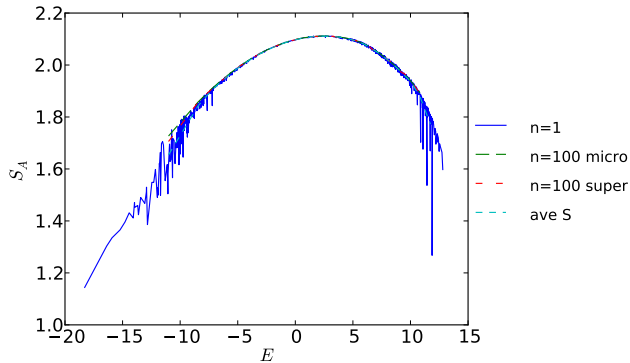}\
(b)\includegraphics[width=\hsize]{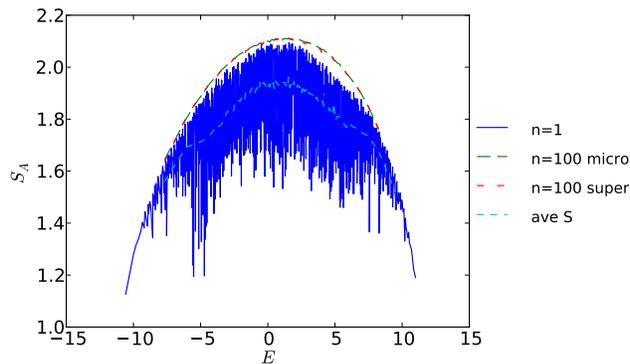}
\caption{ 
(a) The entanglement entropy $S_{Ent}(A,B)$ (Eq. \ref{eq:S_Ent}), for a non-integrable system of $6$ hard core bosons on $27$ sites where $A$ has $4$
sites and $B$ has $23$. The
blue curve $n=1$ is obtained from the entanglement entropy of individual
energy eigenvectors. The dashed green curve is the microcanonical entropy
obtained by averaging the density matrix over $n=100$ neighboring
eigenvectors. The dashed red curve is obtained from the entanglement
entropy of a wave function that is a random superposition of 100 neighboring levels.
The light blue dashed curve is the curve $n=1$ smoothed over
100 neighboring energy levels. 
(b) The same as in (a) but for an integrable system $t'=V'=0$. 
}
\label{fig:EntCompHCB}
\end{center}
\end{figure}

We study systems that  have been previously shown to give rise to
energy eigenstate thermalization.  Following some of these previous
studies~\cite{SantosRigol,RigolBreakdown,RigolQuantumQuench} we consider
spinless fermions and hard core bosons on a one dimensional lattice
with nearest neighbor (NN) and next nearest neighbor (NNN) hopping,
and interaction. The hopping strengths for NN and NNN are $t$ and $t'$,
likewise we denote the interaction strengths $V$ and $V'$. We study
these systems for parameters where they are known to obey ETH and for
integrable parameters where they do not. 
Following previous work, we use periodic boundary conditions in the subspace
with wavevector $k=1$, rather than $k=0$ to avoid a parity symmetry.
Throughout this work we set
the energy scale to have $t=V=\hbar=1$.

For both the fermions and hard core boson cases, $t'=V'=0.96$ which is
non-integrable and should obey the ETH. We calculate the entanglement
entropy for each energy eigenstate for a range of different bath and
system sizes. Our largest size was $N=27$ lattice sites with $N_p = 6$
particles. 
We trace over the bath sites and calculate the entanglement
entropy Eq. \ref{eq:S_Ent}, and do this as a function of the number
of lattice sites of $A$, denoted $m$.  For these cases, we plotted
the entanglement entropy as a function of $m$. As
is well known~\cite{BennettBernsteinPopescuSchumacher} $S_{ent}(m) =
S_{ent}(N-m)$ so for $m > N/2$ the entanglement entropy must go back down
to zero. As we display in the supplementary information in the case of hard core bosons,
it is very close to linear behavior over a substantial range of $m$, for the non-integrable
case, as is expected due to the extensive nature of entropy. 

However the behavior for integrable systems, $t'=V'=0$, 
is more erratic and the linearity depends much more strongly on the eigenvector.
Similar results were also obtained for the fermion model as well as is shown in the
supplementary material.

We directly checked to see if Eq.  \ref{eq:rho_micro=rho} is supported numerically.
We calculated the left hand side of this equation for different values of $m$
and because of the linearity of the entanglement entropy with $m$,
the results are insensitive to this value. In Fig. \ref{fig:EntCompHCB} it is shown for $m=4$.
We also calculate the microcanonical reduced density matrix by taking the partial trace of Eq. \ref{eq:rhomicro}.
We average the reduced density matrix over $100$ neighboring
eigenvectors, and use that to calculate the entanglement entropy as done
on the right hand side of Eq. \ref{eq:rho_micro=rho}.  The result for $6$
hard code bosons on $27$ sites is shown in Fig. \ref{fig:EntCompHCB}(a).
The blue line is no microcanonical averaging ($n=1$) and the green
dashed curve is the microcanonical average over $n=100$ neighbors.
The two curves are very close to each other in most of the range where the
density of states is high and one has eigenstate thermalization. The light
blue dashed curve takes the entanglement entropy with no microcanonical
averaging and averages it over $100$ neighboring energy levels. The
dashed red line will be discussed below.

\begin{figure}[thp]
\begin{center}
\includegraphics[width=\hsize]{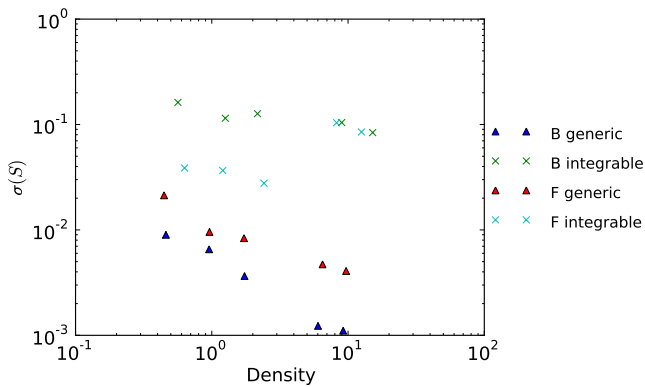}
\caption{ 
The fluctuations in the entanglement entropy for both non-integrable (generic) and integrable
systems of $6$ hard core bosons (B) but with a different number of lattice sites, and the same
quantity for the spinless fermion model (F). The
standard deviation of the entanglement entropy $S_{ent}(A,B)$ is plotted
against the density of states. This is done around the point of maximum
entanglement entropy.
}
\label{fig:EntDiff}
\end{center}
\end{figure}

In contrast, Fig. \ref{fig:EntCompHCB}(b), for the integrable case
$t'=V'=0$, one sees that the fluctuations are much larger and that the
microcanonical average is much further from the entanglement entropy for
individual eigenstates. 
Similar results are seen in the one dimensional spinless fermion system. 
To quantify this difference further, we examine
the standard deviation of fluctuation of the entanglement entropy for
both the integrable and non-integrable cases.

We calculated
the standard deviation $\sigma(S)$ of the entanglement entropy $S_{ent}(A,B)$ around the maximum
of the curves in Fig. \ref{fig:EntCompHCB} for different lattice
sizes, 16, 18, 20, 25, and 27, all with 6 particles in Fig. \ref{fig:EntDiff}. $\sigma(S)$ is computed
over $100$ neighboring energy levels, and this is shown as a function of
the density of states on a log-log plot for both the hard core
boson models (B) and spinless fermions (F), and for both the non-integrable
and integrable cases. As is apparent, the integrable fluctuations are
substantially higher. This is also the case for observables as has been
shown previously~\cite{RigolDunjkoOlshanii}. The behavior as a function
of the density of states and system size is qualitatively different. The
fluctuations are diminishing much more rapidly in the non-integrable
case. This is similar to slower diminution of fluctuations of observables
predicted for integrable versus non-integrable systems, where in the
former case we expect a power law decrease as a function of the number
of degrees of freedom and in the latter case it is predicted to be an
exponential~\cite{DeutschQuantStatMech}.

The above studies provide excellent numerical evidence for the
equivalence of the thermodynamic entropy of non-integrable systems
with the entanglement entropy when the wavefunction is in an energy
eigenstate. We also studied two dimensional hard core bosons and fermions
and reached the same conclusion (see supplementary materials).  Now we
ask what happens with more general initial states. If one starts the
system off in a state with fluctuations in energy in a window,
then we can ask how this evolves for long times. We expect thermalization
of observables, and that the entropy should also be the thermodynamic
entropy. We can test to see if this is the case numerically.

Over long times, the energy eigenvector components of non-integrable
wave functions will have random phases~\cite{Srednicki99}. Therefore we test this case to
see if the entanglement entropy still looks like the microcanonical
entanglement entropy.  In  Fig. \ref{fig:EntCompHCB}(a) the dashed
red line computes the entanglement entropy at each energy by using
wavefunctions that are the superposition of $100$ neighboring energy
eigenvectors with Gaussian amplitudes and random phases. As can be seen,
this matches the predictions of the microcanonical entanglement entropy
very well, and the same conclusion is also reached in the
fermion case. The closeness of the reduced density matrix to the canonical
result is expected for general initial conditions where many
energy eigenstates are present~\cite{LindenPopescuShortWinter}. Our
above analysis shows that for generic systems, when the system is put
in an energy eigenstate, we still obtain the microcanonical entropy, which
is crucial, as otherwise thermodynamics would fail in this important case.

Therefore the equivalence of thermodynamic and the entanglement entropy
for long times should be correct, starting from a wide range of initial
states.

We can now answer the questions that we originally posed. Although the
statistical mechanical entropy is most simply understood from the density
of states, for a generic Hamiltonian, the self-entanglement of a single
energy eigenstate can be used to obtain the same result as well. Knowledge
of neighboring levels is not necessary as this entropy as we have seen,
is a property of a single energy energy eigenvector.  Going back to
our earlier comparison with classical physics, we can now answer our
original question. A complete knowledge of the system's quantum state,
does not affect its behavior with respect to macroscopic measurements of
the entropy. An experimenter who measured a systems's state precisely,
will in subsequent measurements obtain the same answers as someone
who is not made privy to this information, despite the latter describing the system
in a mixed state. For long times, a system in a pure state and one in
an statistical ensemble have identical thermodynamic entropies in the
limit of large systems. This is because the entropy in such experiments
measures the system's self-entanglement, not the lack of knowledge of it.

We thank M. Rigol and M.R. Peterson for useful discussions.
Haibin Li is supported by State Scholarship Fund(No. 2010833088)

\end{document}

\title{Supplementary materials to: The microscopic origin of thermodynamic entropy in isolated systems}

\author{J. M. Deutsch}
\affiliation{Department of Physics, University of California, Santa Cruz CA 95064}
\author{Haibin Li}
\affiliation{Department of Physics, University of California, Santa Cruz CA 95064 and 
Department of Applied Physics, Zhejiang University of Technology, Hangzhou
310023, P. R. China}
\author{Auditya Sharma}
\affiliation{International Institute of Physics, Federal University of Rio Grande do Norte, Brazil}
\maketitle

We start by considering two kinds of one dimensional  lattice systems,
hard core bosons (HCB) and spinless fermions (SF). We choose periodic boundary conditions and
add both nearest-neighbor (NN) and next-nearest-neighbor (NNN) hopping and
interactions. Following the notation of Santos and Rigol~\cite{SantosRigol}, the Hamiltonian for bosons $H_B$ and for fermions $H_F$ 
are
\begin{equation}
H_B =  \sum_{i=1}^{L} \left[ -t \left( b_i^{\dagger} b_{i+1} + h.c. \right) +V n_i^b   n_{i+1}^b \right.
- \left. t' \left( b_i^{\dagger} b_{i+2} + h.c. \right) +V' n_{i}^b  n_{i+2}^b \right],
\label{bosonHam}
\end{equation}
and 
\begin{equation}
H_F = \sum_{i=1}^{L} \left[ -t \left( f_i^{\dagger} f_{i+1} + h.c. \right) +V n_i^f   n_{i+1}^f \right.
- \left. t' \left( f_i^{\dagger} f_{i+2} + h.c. \right) +V' n_{i}^f  n_{i+2}^f 
\right].
\label{fermionHam}
\end{equation}

Here, $L$ is the length of the chain, $b_i$ and $b_i^{\dagger}$
are bosonic and  $f_i$ and $f_i^{\dagger}$ are the fermionic annihilation and creation
operators for site $i$. $n_i^b= b_i^{\dagger} b_i$ is the boson and  $n_i^f= f_i^{\dagger} f_i$
is the fermion local density operator. Hardcore bosons cannot occupy the same
site, and the operators commute on different sites. The NN and NNN hopping 
strengths are respectively $t$  and $t'$. The interaction strengths are $V$ and $V'$ respectively. 
We take $\hbar = t =V=1$.

Because of translational invariance and particle number conservation, the
Hamiltonian divides into independent blocks each with fixed total momentum $k$.
As previously pointed out~\cite{SantosRigol}, to avoid degeneracy associated 
with parity symmetry at $k=0$, we take $k=1$ and checked that the energy levels
obey Wigner-Dyson statistics.

\begin{figure}[htp]
\begin{center}
(a)\includegraphics[width=.4\vsize]{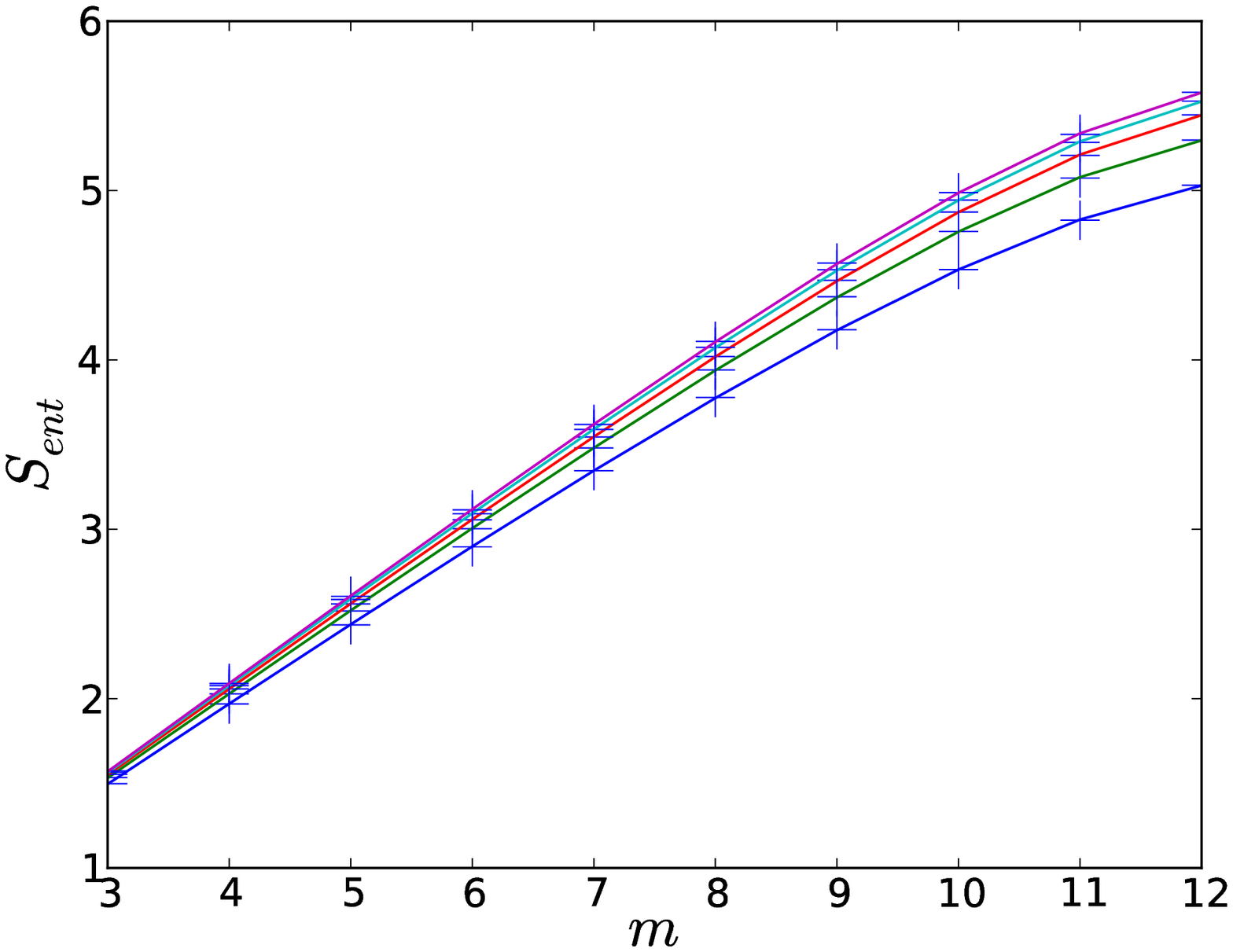}\
(b)\includegraphics[width=.4\vsize]{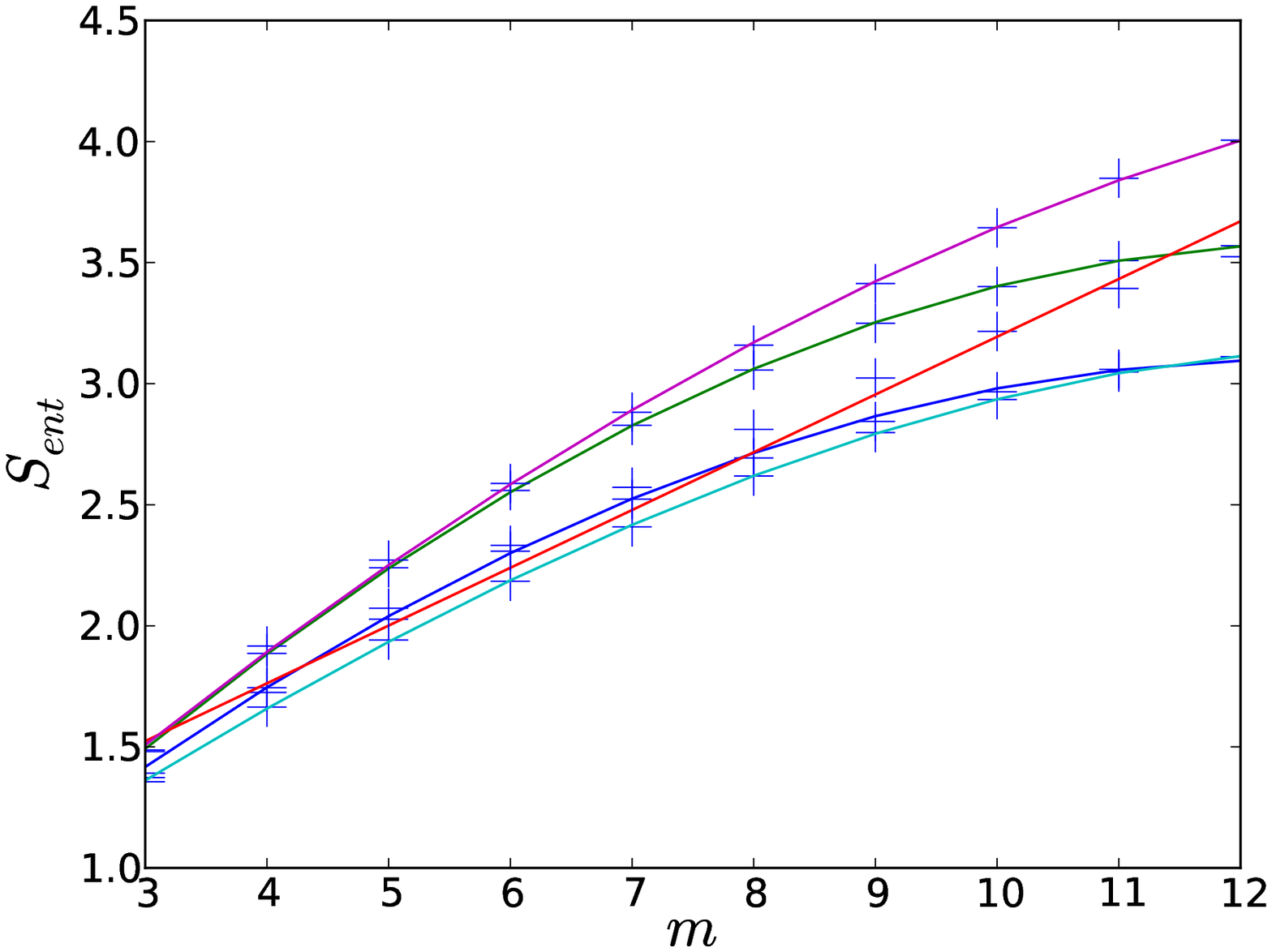}
\caption{ 
The entanglement entropy for single energy eigenstates of hard core bosons shown for a number of
different levels in the non-integrable case. The levels were chosen to be evenly spaces and are 548 (lowest), 1096, 1644, 2192, 2740 (highest)
out of a total of 10966 states in the $k=1$ sector 
(a) The non-integrable case, 
for  $t'=V'=0.96$. The fit was to a straight line plus and exponential correction and the
standard deviation of the error for each curve is less than $3.4\times 10^{-3}$ for all of these levels.
(b) The integrable case $t'=V'= 0$.
}
\label{fig:SentHCB}
\end{center}
\end{figure}

\begin{figure}[htp]
\begin{center}
(a)\includegraphics[width=.4\vsize]{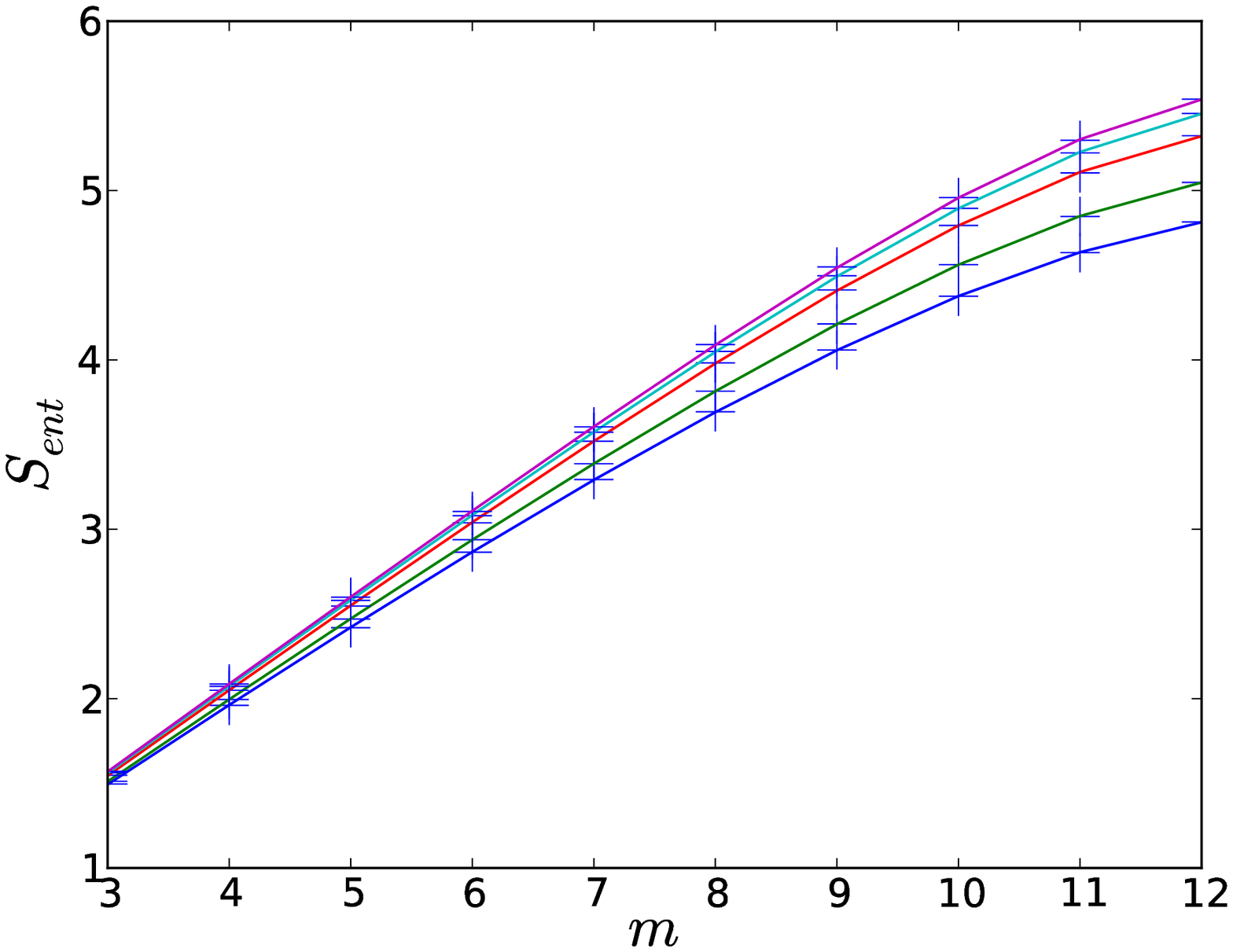}\
(b)\includegraphics[width=.4\vsize]{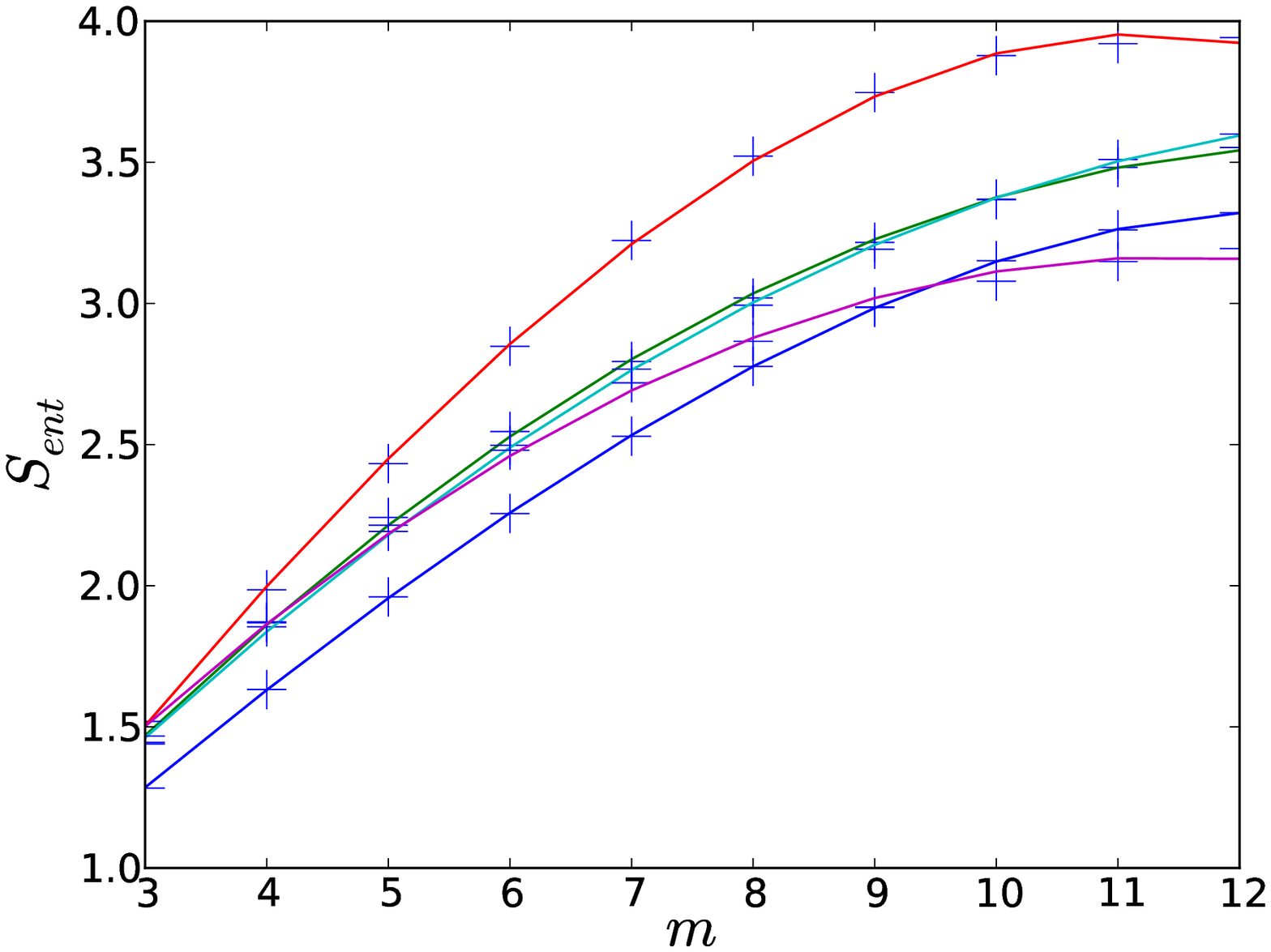}
\caption{ 
The entanglement entropy for single energy eigenstates of spinless fermions shown for a number of
different levels in the non-integrable case. The levels were chosen to be evenly spaces and are 548 (lowest), 1096, 1644, 2192, 2740 (highest)
out of a total of 10966 states in the $k=1$ sector 
(a) The non-integrable case, 
for  $t'=V'=0.96$. The fit was to a straight line plus and exponential correction and the
standard deviation of the error for each curve is less than $3.1\times 10^{-3}$ for all of these levels.
(b) The integrable case $t'=V'= 0$.
}
\label{fig:SentF}
\end{center}
\end{figure}

Using exact diagonalization, we computed the eigenvalues and eigenvectors of the
Hamiltonian in the $k=1$ sector. We studied the cases $V' = t' = .96$, and $V' = t' = 0$. 
The HCB results for the entanglement entropy $S_{ent}(m)$ are shown in Fig. \ref{fig:SentHCB}(a) for the non-integrable case $V' = t' = .96$
and the in integrable case Fig. \ref{fig:SentHCB}(b) $V' = t' = 0$. Here the
number of lattice sites $N=27$, and the number of particles $N_p = 6$.
The five plots sample different eigenvectors and are evenly
spaced as a function of the eigenvector index.  The plots start at $m=3$
because for $m$ smaller than a correlation length, we do not expect
to obtain the correct entropy.  The corrections 
to linearity are well fitted with a small exponential
term. For the integrable case (b), the
behavior is much more erratic and less straight than in the non-integrable
case(a). Fig. \ref{fig:SentF} we display the analagous results for spinless
fermions with the same parameters.

\begin{figure}[htp]
\begin{center}
(a)\includegraphics[height=0.4\vsize]{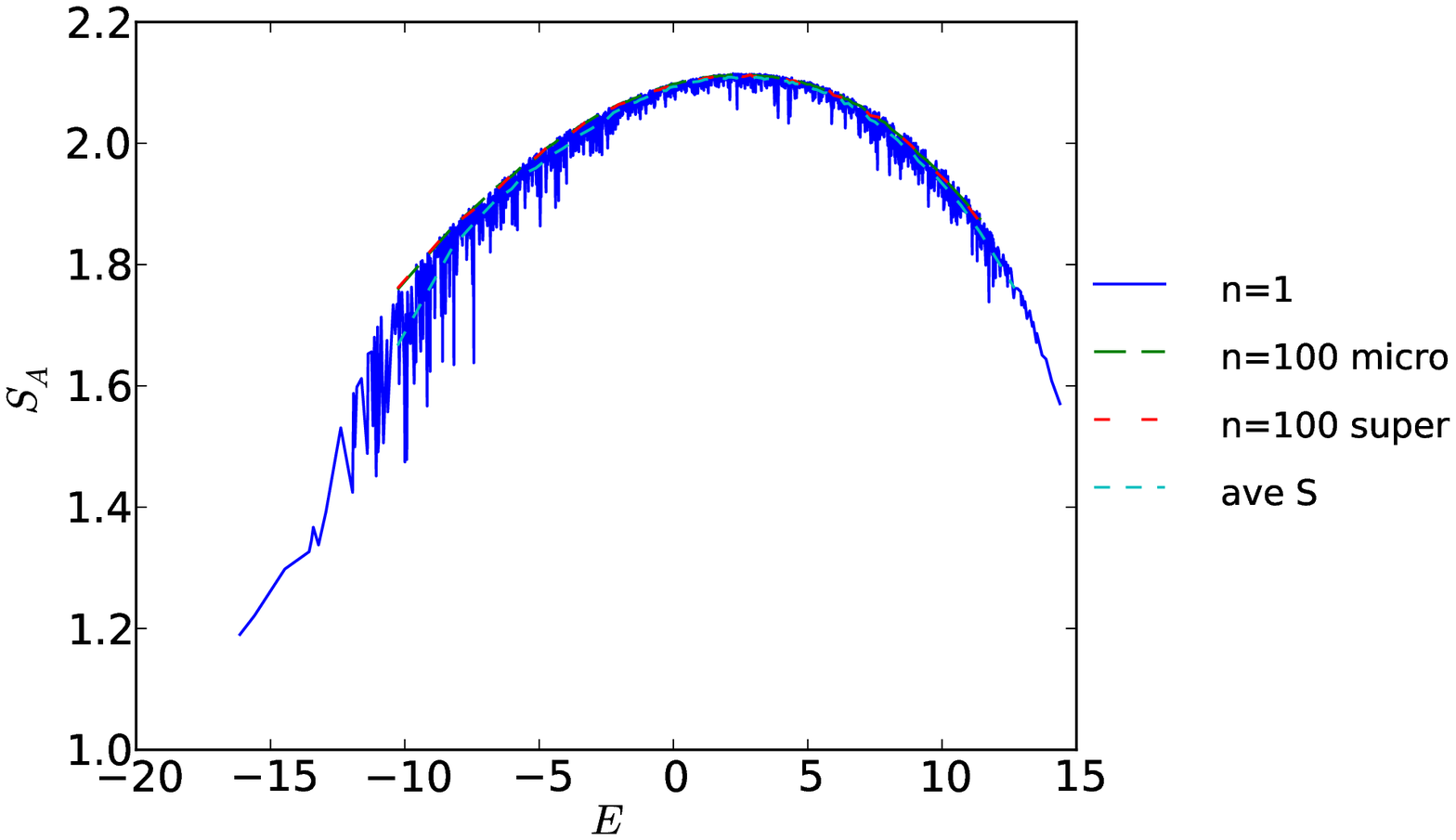}\
(b)\includegraphics[height=0.4\vsize]{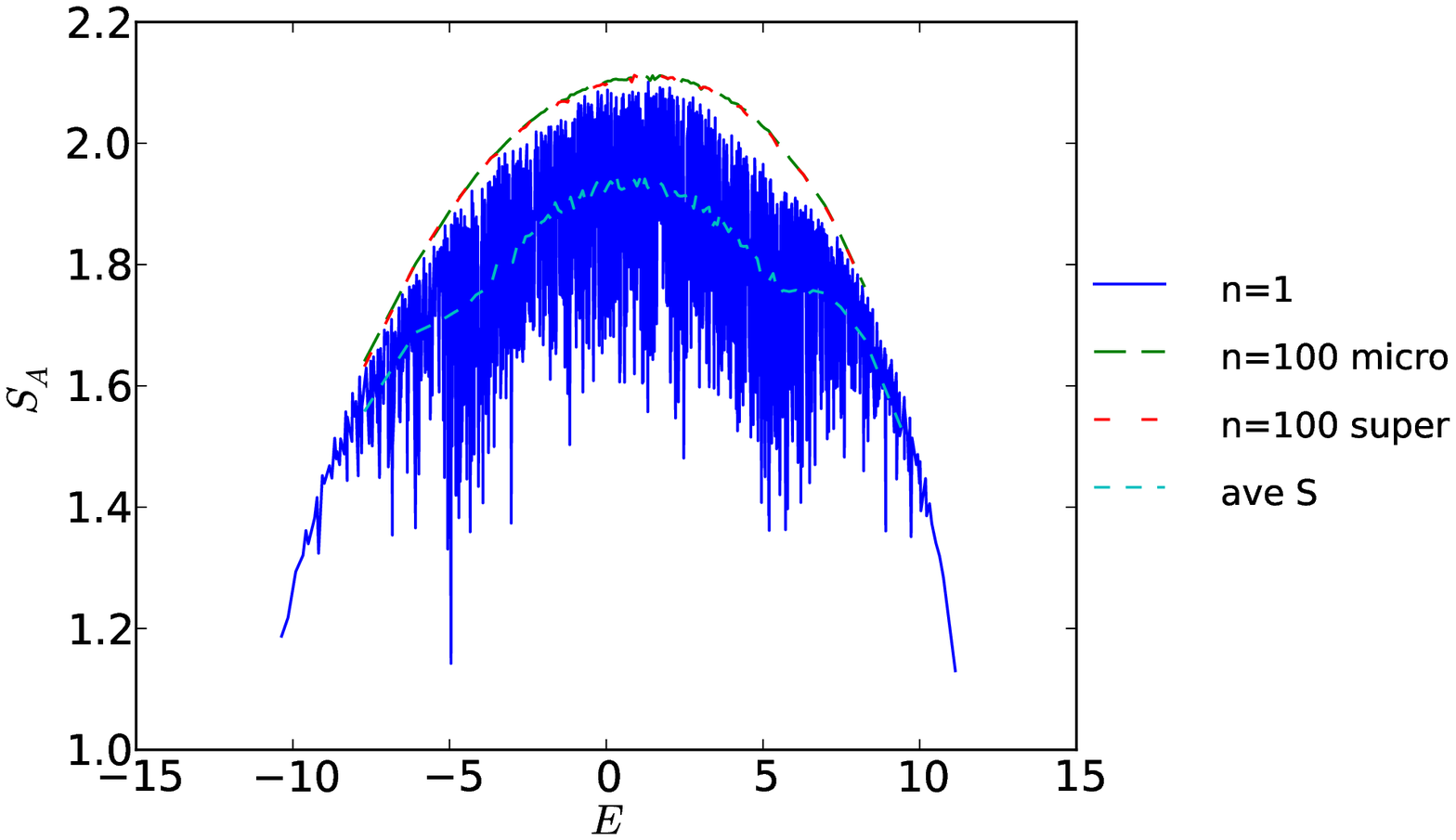}
\caption{ 
(a) The entanglement entropy $S_{Ent}(A,B)$, for a non-integrable system of $6$ spinless fermions on $27$ sites where $A$ has $4$
sites and $B$ has $23$. The
blue curve $n=1$ is obtained from the entanglement entropy of individual
energy eigenvectors. The dashed green curve is the microcanonical entropy
obtained by averaging the density matrix over $n=100$ neighboring
eigenvectors. The dashed red curve is obtained from the entanglement
entropy of a wave function that is a random superposition of 100 neighboring levels.
The light blue dashed curve is the curve $n=1$ smoothed over
100 neighboring energy levels. 
(b) The same as in (a) but for an integrable system $t'=V'=0$. 
}
\label{fig:EntCompSF}
\end{center}
\end{figure}

We show the same quantities as shown in Fig. 2 of the letter for spinless
fermions in Fig. \ref{fig:EntCompSF}, with the same parameters as used for the hard core bosons.

\begin{figure}[htp]
\begin{center}
(a)\includegraphics[width=0.4\vsize]{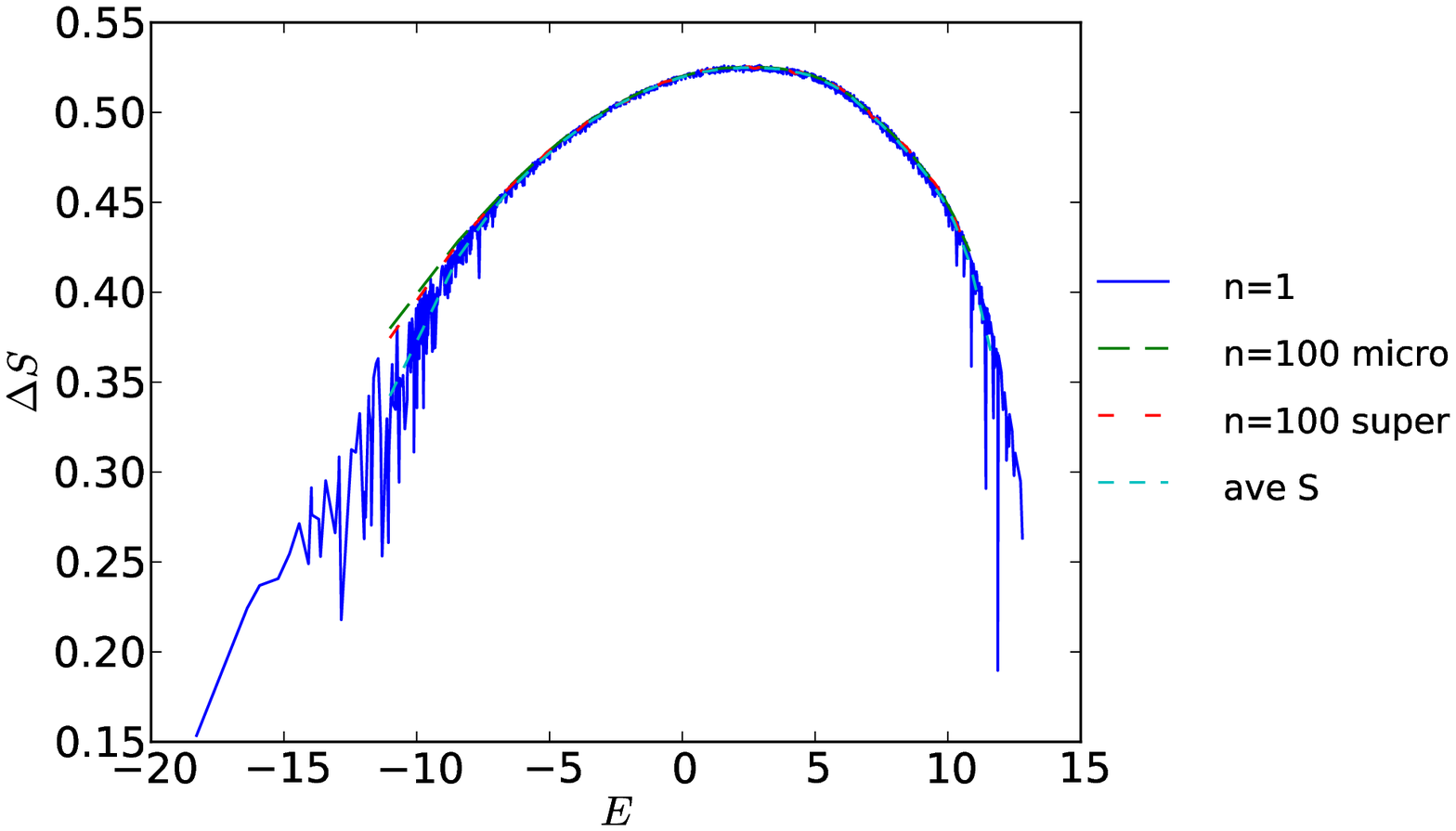}\
(b)\includegraphics[width=0.4\vsize]{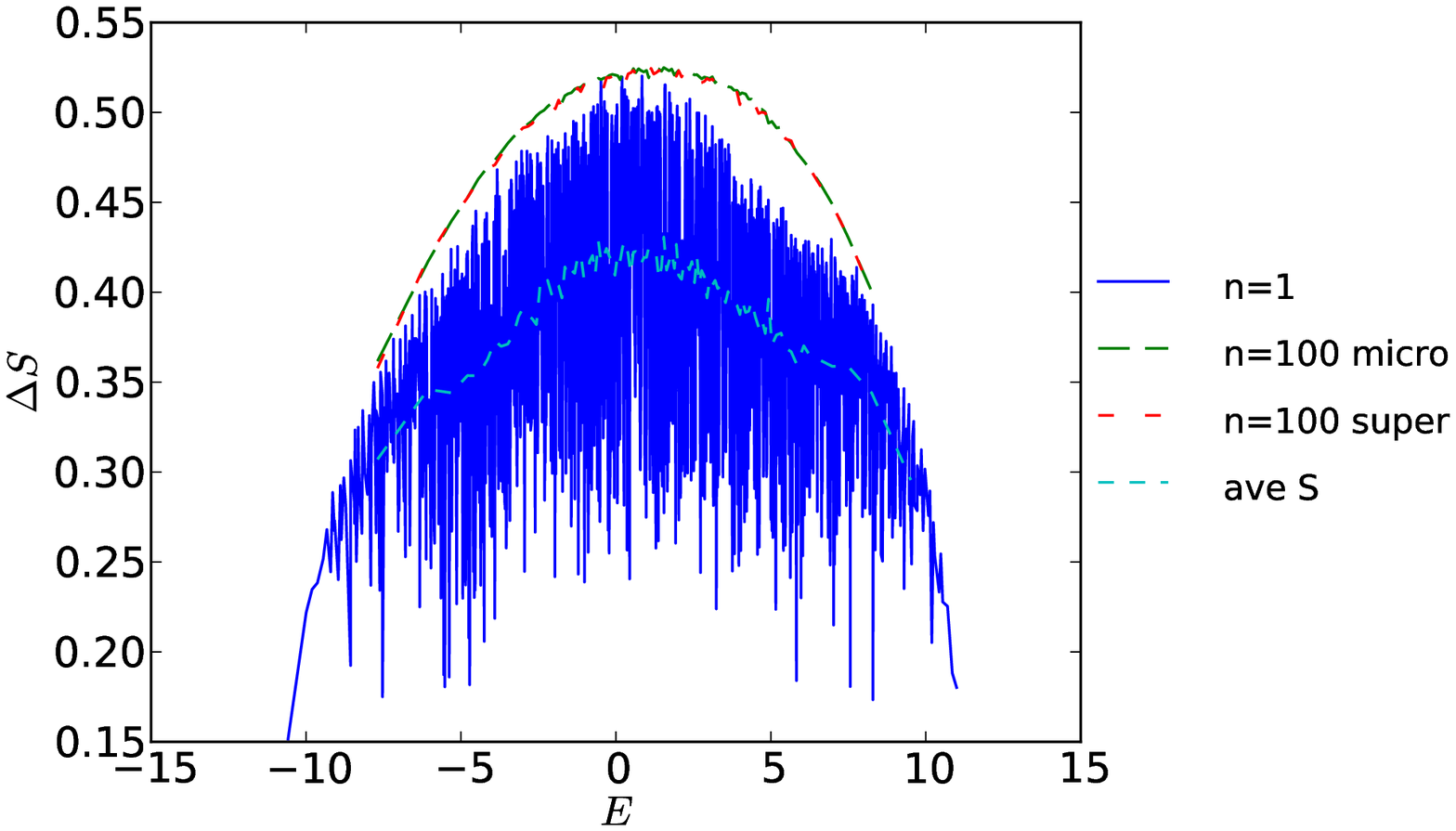}
\caption{ 
(a) The entanglement entropy difference for a non-integrable system of $6$ bosons on $27$ sites for the same
system as Fig. 1 of the accompanying letter. 
Here $\Delta S \equiv S_{ent}(4)-S_{ent}(3)$.
The blue curve $n=1$ is obtained from the entanglement entropy of individual
energy eigenvectors. The dashed green curve is the microcanonical entropy
obtained by averaging the density matrix over $n=100$ neighboring
eigenvectors. The dashed red curve is obtained from the entanglement
entropy of a wave function that is a random superposition of 100 neighboring levels.
The light blue dashed curve is the curve $n=1$ smoothed over
100 neighboring energy levels. 
(b) The same as in (a) but for an integrable system $t'=V'=0$. 
}
\label{fig:EntCompDiffHCB}
\end{center}
\end{figure}

Because of the strong linearity in the entanglement entropy as a function of $m$, 
we can take the difference $\Delta S \equiv S_{ent}(4)-S_{ent}(3)$ to be an estimate for the entropy
of the system per lattice site. The results are very similar to those
obtained using $\Delta S \equiv S_{ent}(5)-S_{ent}(4)$ instead. The results for
hard core bosons are displayed in Fig. \ref{fig:EntCompDiffHCB}.
This difference displays the same features as the full entanglement entropy and
still shows very good agreement with the microcanonical results in the case
of a non-integrable system, but does not agree well for the integrable case.

We next consider the same quantities for a two dimensional square lattice of
hard core bosons with periodic boundary condition. We chose the total momentum
sector to be zero. Hard core bosons in two dimensions have also been shown to satisfy energy eigenstate
thermalization~\cite{RigolDunjkoOlshanii}. We consider 6 hard core bosons on
a 5 by 5 lattice. We choose repulsive nearest neighbor interactions $V = 0.1$
and a nearest neighbor hopping strength $t=1$. As above we chose  $\Delta S \equiv S_{ent}(4)-S_{ent}(3)$

\begin{figure}[htp]
\begin{center}
\includegraphics[width=\hsize]{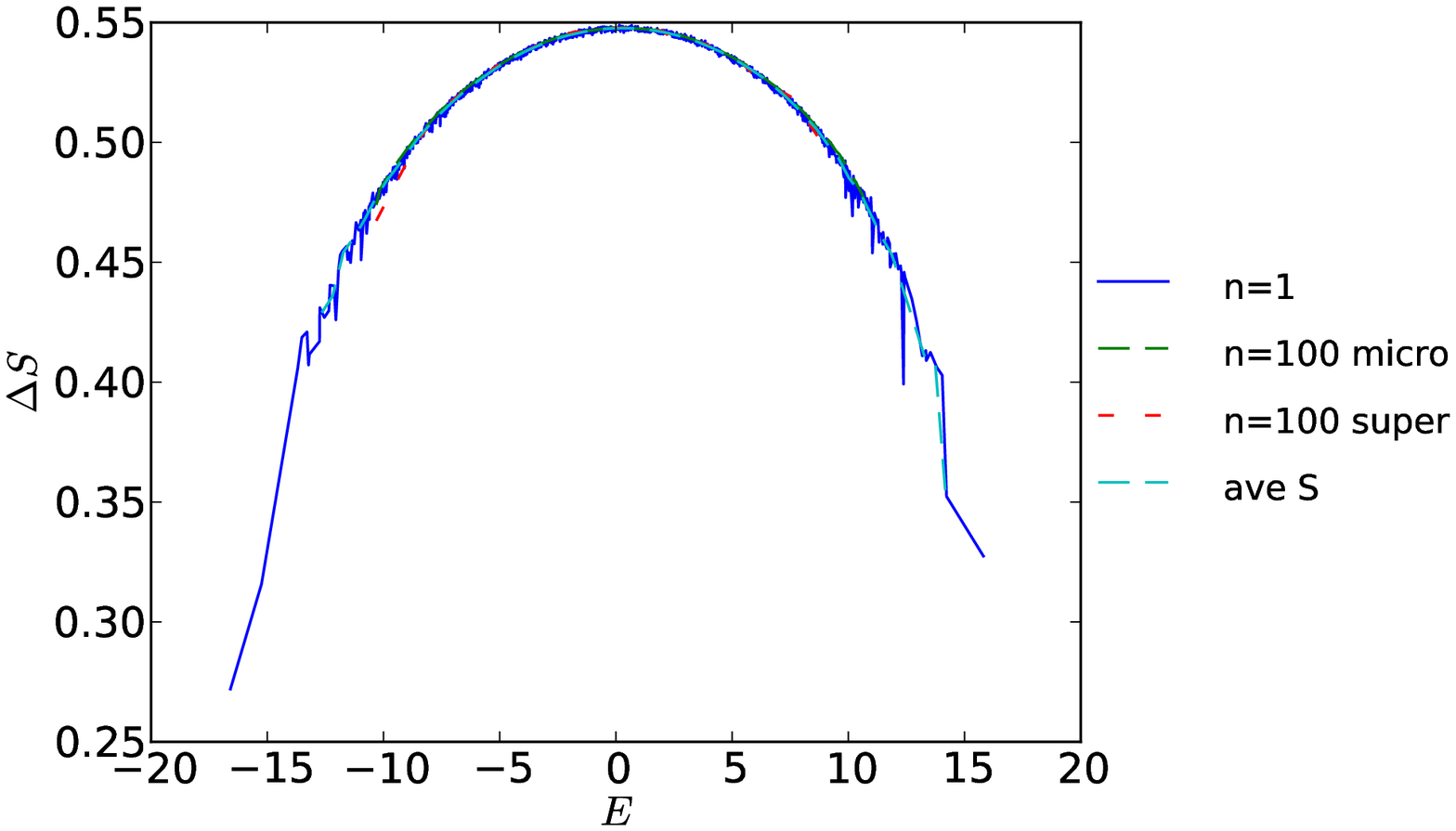}
\caption{ 
The entanglement entropy difference for a non-integrable system of $6$ hard core bosons on a two dimensional  $5\times5$ lattice. 
Here $\Delta S \equiv S_{ent}(4)-S_{ent}(3)$.
The blue curve $n=1$ is obtained from the entanglement entropy of individual
energy eigenvectors. The dashed green curve is the microcanonical entropy
obtained by averaging the density matrix over $n=100$ neighboring
eigenvectors. The dashed red curve is obtained from the entanglement
entropy of a wave function that is a random superposition of 100 neighboring levels.
The light blue dashed curve is the curve $n=1$ smoothed over
100 neighboring energy levels. 
}
\label{fig:EntCompDiff2dHCB}
\end{center}
\end{figure}

Fig.  \ref{fig:EntCompDiff2dHCB} shows the results from exact diagonalization
of this model. As we found in one dimensional non-integrable systems, the
microcanonical entanglement entropy is in very good agreement with the
entanglement entropy.

\newpage

\newpage

\newpage

\newpage

\newpage
